\documentclass[nonacm]{acmart}

\acmConference[SimuRec@RecSys '21]{SimuRec Workshop in conjunction with with the 15th ACM Conference on Recommender Systems}{27th September-01 October, 20218}{Amsterdam, The Netherlands}

\setcopyright{none}
\begin{document}

\title[Doing Data Right]{Doing Data Right: How Lessons Learned Working with Conventional Data should Inform the Future of Synthetic Data for Recommender Systems}

\author{Manel Slokom}
\email{m.slokom@tudelft.nl}
\affiliation{%
   \institution{Delft University of Technology}
  \country{Netherlands}
 }

\author{Martha Larson}
\email{mlarson@science.ru.nl}
\affiliation{%
   \institution{Radboud University}
  \country{Netherlands}
 }

\begin{abstract}

We present a case that the newly emerging field of synthetic data in the area of recommender systems should prioritize `doing data right'. 
We consider this catchphrase to have two aspects: First,  we should not repeat the mistakes of the past, and, second, we should explore the full scope of opportunities presented by synthetic data as we move into the future.
We argue that explicit attention to dataset design and description will help to avoid past mistakes with dataset bias and evaluation.
In order to fully exploit the opportunities of synthetic data, we point out that researchers can investigate new areas such as using data synthesize to support reproducibility by making data open, as well as FAIR, and to push forward our understanding of data minimization.

\end{abstract}


\maketitle
\section{Introduction}
The era of big data has seen impressive examples of how knowledge and value can be created using data.
It has also seen sobering reminders of how easy it is to `do data wrong', causing unintended outcomes and often outright harm to the people~\cite{DataJustice}.
In this position paper, we point out that we are at the beginning of a new era of synthetic data and that we should take this beginning as an opportunity to `do data right'.

\emph{Synthetic data} is data that is created to serve in the place of original data, which is directly collected or captured.
Synthetic data makes it possible to carry out analyses or develop new algorithms on data that would be otherwise too sensitive to retain, share, or release.
Synthetic data can also be used to augment an existing dataset to improve the performance of algorithms.
Data that is only partially synthesized is referred to as \emph{semi-synthetic data}.
The goal of this paper is remind researchers that we must not repeat mistakes that have been made in the past and must also ensure that as our research moves forward, we take advantage of the full scope of the opportunities presented by synthetic data.

Our focus is on recommender system data, which takes the form of a user-item matrix $\mathcal{R}$ with $U= \{1, .., N\}$ users and $ I= \{1, .., M\}$ items. 
If a user has interacted with an item, the corresponding matrix cell contains a 1.
Interactions can include clicks, views and purchases, which implicitly express a preference of a user.
Today, recommender system research focuses on such \emph{implicit data}, since \emph{explicit data}, which consists of ratings explicitly expressing user preferences, is harder to come by.
Recommender system data differs from many datasets of the big data era in that it is highly sparse and characterized by long-tail distributions.
Specifically, we have active users who watch/consume/click many items and non-active users (including cold start users) who attempt to watch/consume/click very few items.
Similarly for items, we have popular items that are consumed by many users and non popular items are consumed by few users.

The special characteristics of recommender system datasets make them challenging to synthesize, and research has just begun in this direction. 
This means that the time is right to avoid the pitfalls already encountered with conventional data.
Further motivation past failure in `doing data right' with regards to user privacy, which is still fresh in the minds of recommender system researchers.
Memorably, in 2010, NetFlix Prize competition was discontinued after it was demonstrated that the data that was released to allow the competitors to develop recommender systems could be deanonymized, revealing the identity of individual users~\cite{narayanan2008robust}.
Flashing forward, the NetFlix Prize debacle has inspired research on synthetic data.
Via the experience of our own research we see two directions emerging. 
First, research on using synthetic or semi-synthetic data to replace captured data in competitions~\cite{slokom2019data}.
Second, research on ensuring that synthetic or semi-synthetic data that is derived from data originally collected from users does not leak sensitive information on those users~\cite{slokom2021PerBlur}.

The paper is structured as follows.
First, we look at two areas, going beyond privacy, where past research in recommender systems arguably failed in `doing data right' when working with conventional: bias and evaluation.
We discuss how research in synthetic data can grab the chance of not repeating past mistakes.
Then, we discuss two opportunities that are opened by synthetic data, which are not offered by conventional data: open data and data minimization.
We present remarks on how the recommender system community can build on these opportunities.
Our paper closes with a short summary and an outlook.

\section{Addressing Past Mistakes}
The era of big data has been driven by the idea that more data will automatically give rise to more reliable analysis and better systems.
In recent years, however, machine learning researchers have initiated a more systematic approach to data in which the quality, not just quantity, of data is central.
These efforts are well represented by the initiative of \emph{datasheets for datasets}~\cite{gebru2018datasheets}.
In a nutshell, this initiative proposes that every dataset is described by a datasheet with a standardized format that documents: the motivation (why a dataset is created), creation (how the dataset is created), composition (what information it contains), intended uses (what tasks it should (not) be used for), data distribution (what are the properties of the dataset).
In this section, we look at past cases of `doing data wrong' related to data bias and to evaluation.
We comment on how understanding, documenting, as well as explicitly designing, the characteristics of data is currently offering course correction for research practices and also on how work on synthetic data can be steered so that the same problems that we have confronted while working with conventional data do not arise anew.

\subsection{Bias Mitigation}

In its early days, the recommender systems community did not considered issues of bias and fairness.
Thankfully, recent work has started to illuminate these issues. Here, we provide a brief summary.
Discrimination and unfairness in recommender systems can originate from different sources: 
First, \textit{input bias}~\cite{tsintzou2018bias, lin2019crank} that users exhibit in the input data. 
In~\cite{lin2019crank}, the authors studied how different collaborative filtering algorithms propagate bias existing in the input data and its impact on users.
In~\cite{Cool2018ekstrand}, the authors evaluated the ability of recommender system algorithms to produce equal utility for users of different demographic groups.
A set of results showed a statistically significant differences in effectiveness between users' gender and age groups.
Second, \textit{algorithmic bias}~\cite{tsintzou2018bias, mansoury2019bias} examines the effectiveness of recommendation algorithms in capturing different users' interests across item categories.
For example, popularity bias, where the recommender gives higher accuracy scores to algorithms that favor popular items irrespective of their ability to meet user needs. 
In~\cite{edizel2019fairecsys}, the authors proposed FaiRecSys, an algorithm that mitigates algorithmic bias by post-processing the recommendation matrix with minimum impact on the accuracy of recommendations provided to the end-users.
Third, \textit{evaluation metric error and bias}~\cite{tian2018monte} simulates the recommender data generation and evaluation processes to quantify how erroneous current evaluation practices are. 
In~\cite{yao2021measuring}, the authors proposed a simulation framework for measuring the impact of a recommender system under different types of user behavior. 
The framework goes beyond one-step recommendation and incorporates the interaction between user preferences and system effects, to better understand recommender system biases over time.

Biased data, biased algorithm and a biased metric will have an impact on all users with different degrees, which leads to discrimination, unfairness and harm.
Data synthesis is an important approach to mitigate bias.
Synthesized data can potentially support recommender systems' experimentation, tuning, validation and performance prediction.
When synthesizing data, there are some points that we attempt to achieve or test.
For instance, the (semi-)synthesized data can be used to mitigate bias~\cite{krishnan2014methodology,huang2020keeping}, improve consumer-provider fairness~\cite{li2021user,boratto2020interplay}, data augmentation~\cite{belletti2019scaling}.

We argue that although data synthesis is helpful to address bias, alone it is not enough.
It is critical that the design decisions that were made when creating a synthesized dataset are well motivated, and made explicit, and also that they are well documented.
In this way, future researchers can understand how bias was handled and assure themselves that new forms of bias were not introduced during the synthesis process.
With explicit design and careful documentation, we can learn, understand, and explain where things have gone wrong and ideally be able to work toward redressing problem i.e., harms and preventing further problems.
The goal of datasheets for datasets is to provide more transparency, accountability and control in the machine learning and recommender system communities.
Moving forward it is crucial that datasheets are also crated for synthetic data.

\subsection{Reliable Evaluation}
In its early days, the recommender systems community did not fully appreciate the importance of systematic evaluation.
Arguably, it was~\cite{saidbellogin2014} that awakened researchers to the importance of completely controlling the dimensions of an evaluation in order to achieve a fair comparison.
The first dimension mentioned by~\cite{saidbellogin2014} is data.
In recent years, the community has made strides in evaluation practices and reproducibility, see~\cite{bellogin2021improving}, which contains a section documenting the effort.
We point out that a datasheets approach to synthetic data, will ensure that synthetic data will be used appropriately for evaluation from the start and invalid comparisons between datasets will be avoided.
Another dimension mentioned by~\cite{saidbellogin2014} is evaluation strategies. Here, we dive deeper to discuss why careful attention must be paid to evaluation strategies for synthetic or semi-synthetic data.

In the machine learning literature, the quality of synthetic data is often evaluated using machine learning performance.
Such an evaluation involves comparing the performance metrics of predictive models trained on synthetic and on real data (called as model compatibility).
This performance of a machine learning models trained and tested on real and or synthetic data is compared based on different scenarios~\cite{heyburn2018machine, jordon2018measuring, fekri2020generating}: 
Train on Real and Test on Synthetic data ($\mathcal{TRTS}$)
Train on Synthetic and Test on Real ($\mathcal{TSTR}$), Train on Real, Test on Real ($\mathcal{TRTR}$) and Train on Synthetic, Test on Synthetic ($\mathcal{TSTS}$), and lastly trained and tested on a mixture of real and synthetic data ($\mathcal{TMTM}$).
In principle, these scenarios are transferable to the evaluation of synthetic data in recommender systems.
However, it is important to consider whether $\mathcal{TRTS}$ and
$\mathcal{TSTR}$ 
actually yield meaningful information about how useful synthetic data is for recommendation.
The reason is that, if the synthetic data provides synthetic users, then users in the training set (or test set) are different from those in the test set (respectively training set).

It is critical to develop evaluation frameworks that are suitable for use in evaluating synthetic data in the context of recommender systems.
In other words, evaluation itself must be an object of research.
Here, we cite two directions that could serve as a starting point.
First, relative ranking of a set of algorithms, rather than absolute scores could serve as an important tool.
The relative performance of a set of algorithms trained and tested on the synthetic dataset should be the same as their relative performance when trained and tested on the original dataset~\cite{jordon2018measuring,jordon2018pate,bowen2019comparative,slokom2019data}.
For example, if (semi-)synthetic data is released for use in a data science challenge, this relative ranking would be more important that the absolute scores achieved by the algorithms.
This direction of research is not yet well explored by researchers in recommender system community.
Second, special attention must be paid to ensure that the test set remains comparable when different types of (semi-)synthetic data are compared.
We have proposed on way to address this issue for semi-synthetic data~\cite{slokom2021PerBlur}.

We close this section on evaluation by mentioning the importance of studying data characteristics.
There is an interaction between the exact nature of the data, and the types or recommender system algorithms that perform well on that data.
These aspects have traditionally been understudied by the community, also the situation is hopefully changing in the wake of~\cite{Adomavicius20212Impact,deldjoo2021explaining}.
Because it is straightforward to control the properties of synthetic data, the study of synthetic data opens a whole new world of possibilities for use to understand which algorithms works well with which type of data, and why.
Again, we see that the proper documentation of synthetic data in datasheets is critical for such research to be reproducible and thereby useful.

\section{Paving the Way for Future Research}
Synthetic data is generally intended to take the place of original data.
However, in order to take advantage of the full potential of synthetic data, which must also invest research effort in developing the potential of synthetic data to transcend conventional data, and be used for purposes for which conventional data is not suited.

\subsection{FAIR and Beyond}

FAIR is the combination of different small practices that make the data easier to \textit{find}, easier to \textit{understand}, less likely to be \textit{lost}, and more likely to be \textit{usable} during the project time and years later~\cite{Inau2021Initiatives}.
FAIR principles~\cite{GoFair} are guidelines for data management and stewardship that are valid for both machines and humans:
\textit{Findable}: (meta)data should be discoverable, identifiable and searchable via the assignment of metadata and unique identifiers.
\textit{Accessible}: (meta)data should be available and retrievable with access via authentication and authorisation procedures.
\textit{Interoperable}: (meta)data should be semantically understandable, allowing the broadest possible data exchange i.e., exchange and reuse between researchers, institutions, organisations or countries.
\textit{Reusable}: (meta)data should be sufficiently described, well documented, and shared with the least restrictive licenses, allowing the widest reuse possible.

The FAIR principles can drive forward progress in recommender system research because they can support reproducibility.
However, the FAIR principles do not dictate that the data has to be shared openly~\cite{OpenAIRE},
which is a hindrance to reproducibility.
For instance, \textit{the data can be FAIR but not open}: it is FAIR within the company but it does not open to researchers, scientists and users outside the company.
Data synthesis offers a possibility to make data
FAIRly open without the need to release the original data.
The (semi-)synthetic data could be designed to protect user's sensitive information while still maintaining its value for training recommender systems, which is needed for reproducibility.
We have suggested one approach in~\cite{slokom2021PerBlur}, but this work represents only a beginning.
The (semi-)synthetic data could also be designed to protect information that is important for companies' competitiveness while at the same time preserving the information that is necessary for the data to contain in order for third-parties to be able to have oversight over how companies collect and use the data of users.

\subsection{Data Minimization}
Finally, we discuss the issue of data minimization. 
Article 5(1)(c) of the European Union's General Data Protection Regulation (GDPR) requires that personal data should be limited to only what is necessary to the purposes for which the data is processed~\cite{regulation2018principles}.
Linking back to the discussion of FAIR, we note that in~\cite{Inau2021Initiatives, boeckhout2018fair}, authors suggested that FAIR data and metadata can facilitate compliance with data minimization principle since FAIR principles allow for an assessment of which data to reuse.

Here, we zero in specifically on data minimization for recommender systems.
In~\cite{larson2017towards}, the authors proposed to adopt training data requirements analysis to analyze and evaluate the trade-off between the amount of data that the system requires, and the performance of the system.
In~\cite{krishnaraj2019comparative}, the authors proposed to extend the data minimzation principles advocated in GDPR and studied their effect on recommender systems.
They investigated the effects of reducing the amount of data used to model a recommender system and showed that a substantial amount of data can be dropped without a large impact on the performance.
In~\cite{biega2020operationalizing}, authors pointed to the lack of an homogeneous interpretation of the data minimization principle.
They argued that personalization-based systems do not necessarily need to collect user data, but that they do so to improve the quality of the results.
They found that the performance decrease incurred by data minimization might not be substantial but that it might disparately impact different users.
To support minimization, \cite{biega2020operationalizing} suggested that we need to design new protocols for user-system interaction, a system that does not only focus on providing infinite recommendations while collecting infinite data about its users'.
In other words, we need to propose new learning mechanisms that select necessary data that respect specific minimization requirements while maintaining a good personalized-based recommendation performance.

Synthetic data presents a promising opportunity to understand what data minimizing means for recommender systems.
Minimized datasets can be synthesized with different characteristics and the impact of these characteristics could be studied.
We believe that cold start user profiles could be a good starting point to understand and find the minimal necessary data in a user profile.
Then, recommender system research need to look at how much data is really necessary to accomplish a given recommender system task.
We expect the study of data minimization to move forward the state of the art in recommender systems, but also to make it possible to gain understanding of how the GDPR must be enforced for recommender system data.

Using synthetic data to study data minimization is potentially relevant to oversight beyond the GDPR as well.
Currently, there is growing concern about the manipulative impact of hypertargeting, which infringes on privacy and consumer rights.
Previously, we have proposed the concept of hypotargeting~\cite{larson2019up}, i.e., imposing a constraint on the number of unique recommendation lists that a recommender system can present to its users in a given time window.
Because the number of unique lists remains finite, it becomes feasible to audit the experience that a recommender system is offering to its users.
Such oversight can watch for bias, filter bubbles, and unfair targeting.

\section{Summary and Outlook}
In this position paper, we have described mistakes that have occurred over the history of recommender system research, specifically, neglecting the issue of bias and overlooking the importance of evaluation framework.
We have argued that we must ensure that these mistakes are not repeated as we develop approaches to craete synethetic data for evaluation.
We have also pointed to areas where synthetic data has a special contribution to make in the future, specifically, extending FAIR principles to make data open and also moving forward our understanding of data minimization for recommender systems and how to minimize data appropriately and effectively. 

Throughout we have emphasized the importance of explicitly designing and documenting synthetic datasets, following the idea of datasheets for datasets~\cite{gebru2018datasheets}.
Future research will need to embrace the development of best practices for design, documentation, and evaluation of synthetic data as research areas in their own right.

Recommender system research must also create bridges across disciplines. 
As pointed out by~\cite{gebru2018datasheets}, the risk datasets causing harm can be exacerbated when developers are not domain experts.
Moving forward it is essential to include experts from specific domains, such as health, psychology, and communication science, in synthetic data research.
Further, interdisciplinary collaboration is also necessary with legal experts to understand how synthetic data can best protect privacy, and support data minimization and regulatory oversight.



\bibliographystyle{ACM-Reference-Format}
\bibliography{bib}


\begin{thebibliography}{35}


\ifx \showCODEN    \undefined \def \showCODEN     #1{\unskip}     \fi
\ifx \showDOI      \undefined \def \showDOI       #1{#1}\fi
\ifx \showISBNx    \undefined \def \showISBNx     #1{\unskip}     \fi
\ifx \showISBNxiii \undefined \def \showISBNxiii  #1{\unskip}     \fi
\ifx \showISSN     \undefined \def \showISSN      #1{\unskip}     \fi
\ifx \showLCCN     \undefined \def \showLCCN      #1{\unskip}     \fi
\ifx \shownote     \undefined \def \shownote      #1{#1}          \fi
\ifx \showarticletitle \undefined \def \showarticletitle #1{#1}   \fi
\ifx \showURL      \undefined \def \showURL       {\relax}        \fi
\providecommand\bibfield[2]{#2}
\providecommand\bibinfo[2]{#2}
\providecommand\natexlab[1]{#1}
\providecommand\showeprint[2][]{arXiv:#2}

\bibitem[\protect\citeauthoryear{Adomavicius and Zhang}{Adomavicius and
  Zhang}{2012}]%
        {Adomavicius20212Impact}
\bibfield{author}{\bibinfo{person}{Gediminas Adomavicius} {and}
  \bibinfo{person}{Jingjing Zhang}.} \bibinfo{year}{2012}\natexlab{}.
\newblock \showarticletitle{Impact of Data Characteristics on Recommender
  Systems Performance}.
\newblock \bibinfo{journal}{\emph{ACM Transaction on Management Information
  Systems}} \bibinfo{volume}{3}, \bibinfo{number}{1}, Article
  \bibinfo{articleno}{3} (\bibinfo{date}{April} \bibinfo{year}{2012}),
  \bibinfo{numpages}{17}~pages.
\newblock


\bibitem[\protect\citeauthoryear{Belletti, Lakshmanan, Krichene, Mayoraz, Chen,
  Anderson, Robie, Oguntebi, Shirron, and Bleiwess}{Belletti
  et~al\mbox{.}}{2019}]%
        {belletti2019scaling}
\bibfield{author}{\bibinfo{person}{Francois Belletti}, \bibinfo{person}{Karthik
  Lakshmanan}, \bibinfo{person}{Walid Krichene}, \bibinfo{person}{Nicolas
  Mayoraz}, \bibinfo{person}{Yi-Fan Chen}, \bibinfo{person}{John Anderson},
  \bibinfo{person}{Taylor Robie}, \bibinfo{person}{Tayo Oguntebi},
  \bibinfo{person}{Dan Shirron}, {and} \bibinfo{person}{Amit Bleiwess}.}
  \bibinfo{year}{2019}\natexlab{}.
\newblock \showarticletitle{Scaling Up Collaborative Filtering Data Sets
  through Randomized Fractal Expansions}.
\newblock \bibinfo{journal}{\emph{arXiv preprint arXiv:1905.09874}}
  (\bibinfo{year}{2019}).
\newblock


\bibitem[\protect\citeauthoryear{Bellog{\'\i}n and Said}{Bellog{\'\i}n and
  Said}{2021}]%
        {bellogin2021improving}
\bibfield{author}{\bibinfo{person}{Alejandro Bellog{\'\i}n} {and}
  \bibinfo{person}{Alan Said}.} \bibinfo{year}{2021}\natexlab{}.
\newblock \showarticletitle{Improving Accountability in Recommender Systems
  Research Through Reproducibility}.
\newblock \bibinfo{journal}{\emph{arXiv preprint arXiv:2102.00482}}
  (\bibinfo{year}{2021}).
\newblock


\bibitem[\protect\citeauthoryear{Biega, Potash, Daum{\'e}, Diaz, and
  Finck}{Biega et~al\mbox{.}}{2020}]%
        {biega2020operationalizing}
\bibfield{author}{\bibinfo{person}{Asia~J Biega}, \bibinfo{person}{Peter
  Potash}, \bibinfo{person}{Hal Daum{\'e}}, \bibinfo{person}{Fernando Diaz},
  {and} \bibinfo{person}{Mich{\`e}le Finck}.} \bibinfo{year}{2020}\natexlab{}.
\newblock \showarticletitle{Operationalizing the legal principle of data
  minimization for personalization}. In \bibinfo{booktitle}{\emph{Proceedings
  of the 43rd International ACM SIGIR Conference on Research and Development in
  Information Retrieval}}. \bibinfo{pages}{399--408}.
\newblock


\bibitem[\protect\citeauthoryear{Boeckhout, Zielhuis, and Bredenoord}{Boeckhout
  et~al\mbox{.}}{2018}]%
        {boeckhout2018fair}
\bibfield{author}{\bibinfo{person}{Martin Boeckhout},
  \bibinfo{person}{Gerhard~A Zielhuis}, {and} \bibinfo{person}{Annelien~L
  Bredenoord}.} \bibinfo{year}{2018}\natexlab{}.
\newblock \showarticletitle{The FAIR guiding principles for data stewardship:
  fair enough?}
\newblock \bibinfo{journal}{\emph{European journal of human genetics}}
  \bibinfo{volume}{26}, \bibinfo{number}{7} (\bibinfo{year}{2018}),
  \bibinfo{pages}{931--936}.
\newblock


\bibitem[\protect\citeauthoryear{Boratto, Fenu, and Marras}{Boratto
  et~al\mbox{.}}{2020}]%
        {boratto2020interplay}
\bibfield{author}{\bibinfo{person}{Ludovico Boratto}, \bibinfo{person}{Gianni
  Fenu}, {and} \bibinfo{person}{Mirko Marras}.}
  \bibinfo{year}{2020}\natexlab{}.
\newblock \showarticletitle{Interplay between Upsampling and Regularization for
  Provider Fairness in Recommender Systems}.
\newblock \bibinfo{journal}{\emph{arXiv preprint arXiv:2006.04279}}
  (\bibinfo{year}{2020}).
\newblock


\bibitem[\protect\citeauthoryear{Bowen and Snoke}{Bowen and Snoke}{2019}]%
        {bowen2019comparative}
\bibfield{author}{\bibinfo{person}{Claire~McKay Bowen} {and}
  \bibinfo{person}{Joshua Snoke}.} \bibinfo{year}{2019}\natexlab{}.
\newblock \showarticletitle{Comparative study of differentially private
  synthetic data algorithms from the NIST PSCR differential privacy synthetic
  data challenge}.
\newblock \bibinfo{journal}{\emph{arXiv preprint arXiv:1911.12704}}
  (\bibinfo{year}{2019}).
\newblock


\bibitem[\protect\citeauthoryear{Deldjoo, Bellogin, and Di~Noia}{Deldjoo
  et~al\mbox{.}}{2021}]%
        {deldjoo2021explaining}
\bibfield{author}{\bibinfo{person}{Yashar Deldjoo}, \bibinfo{person}{Alejandro
  Bellogin}, {and} \bibinfo{person}{Tommaso Di~Noia}.}
  \bibinfo{year}{2021}\natexlab{}.
\newblock \showarticletitle{Explaining recommender systems fairness and
  accuracy through the lens of data characteristics}.
\newblock \bibinfo{journal}{\emph{Information Processing \& Management}}
  \bibinfo{volume}{58}, \bibinfo{number}{5} (\bibinfo{year}{2021}),
  \bibinfo{pages}{102662}.
\newblock


\bibitem[\protect\citeauthoryear{Edizel, Bonchi, Hajian, Panisson, and
  Tassa}{Edizel et~al\mbox{.}}{2019}]%
        {edizel2019fairecsys}
\bibfield{author}{\bibinfo{person}{Bora Edizel}, \bibinfo{person}{Francesco
  Bonchi}, \bibinfo{person}{Sara Hajian}, \bibinfo{person}{Andr{\'e} Panisson},
  {and} \bibinfo{person}{Tamir Tassa}.} \bibinfo{year}{2019}\natexlab{}.
\newblock \showarticletitle{FaiRecSys: mitigating algorithmic bias in
  recommender systems}.
\newblock \bibinfo{journal}{\emph{International Journal of Data Science and
  Analytics}} (\bibinfo{year}{2019}), \bibinfo{pages}{1--17}.
\newblock


\bibitem[\protect\citeauthoryear{Ekstrand, Tian, Azpiazu, Ekstrand, Anuyah,
  McNeill, and Pera}{Ekstrand et~al\mbox{.}}{2018}]%
        {Cool2018ekstrand}
\bibfield{author}{\bibinfo{person}{Michael~D. Ekstrand}, \bibinfo{person}{Mucun
  Tian}, \bibinfo{person}{Ion~Madrazo Azpiazu}, \bibinfo{person}{Jennifer~D.
  Ekstrand}, \bibinfo{person}{Oghenemaro Anuyah}, \bibinfo{person}{David
  McNeill}, {and} \bibinfo{person}{Maria~Soledad Pera}.}
  \bibinfo{year}{2018}\natexlab{}.
\newblock \showarticletitle{All The Cool Kids, How Do They Fit In?: Popularity
  and Demographic Biases in Recommender Evaluation and Effectiveness}. In
  \bibinfo{booktitle}{\emph{Proceeding FAT*}}
  \emph{(\bibinfo{series}{Proceedings of Machine Learning Research},
  Vol.~\bibinfo{volume}{81})}. \bibinfo{publisher}{PMLR},
  \bibinfo{pages}{172--186}.
\newblock


\bibitem[\protect\citeauthoryear{Fair}{Fair}{[n.d.]}]%
        {GoFair}
\bibfield{author}{\bibinfo{person}{Go Fair}.}
  \bibinfo{year}{[n.d.]}\natexlab{}.
\newblock \bibinfo{booktitle}{\emph{{FAIR} {P}rinciples}}.
\newblock
\newblock
\shownote{\url{https://www.go-fair.org/fair-principles/}, Online; accessed
  22-August-2021.}


\bibitem[\protect\citeauthoryear{Fekri, Ghosh, and Grolinger}{Fekri
  et~al\mbox{.}}{2020}]%
        {fekri2020generating}
\bibfield{author}{\bibinfo{person}{Mohammad~Navid Fekri},
  \bibinfo{person}{Ananda~Mohon Ghosh}, {and} \bibinfo{person}{Katarina
  Grolinger}.} \bibinfo{year}{2020}\natexlab{}.
\newblock \showarticletitle{Generating energy data for machine learning with
  recurrent generative adversarial networks}.
\newblock \bibinfo{journal}{\emph{Energies}} \bibinfo{volume}{13},
  \bibinfo{number}{1} (\bibinfo{year}{2020}), \bibinfo{pages}{130}.
\newblock


\bibitem[\protect\citeauthoryear{Gebru, Morgenstern, Vecchione, Vaughan,
  Wallach, Daum{\'e}~III, and Crawford}{Gebru et~al\mbox{.}}{2018}]%
        {gebru2018datasheets}
\bibfield{author}{\bibinfo{person}{Timnit Gebru}, \bibinfo{person}{Jamie
  Morgenstern}, \bibinfo{person}{Briana Vecchione},
  \bibinfo{person}{Jennifer~Wortman Vaughan}, \bibinfo{person}{Hanna Wallach},
  \bibinfo{person}{Hal Daum{\'e}~III}, {and} \bibinfo{person}{Kate Crawford}.}
  \bibinfo{year}{2018}\natexlab{}.
\newblock \showarticletitle{Datasheets for datasets}.
\newblock \bibinfo{journal}{\emph{arXiv preprint arXiv:1803.09010}}
  (\bibinfo{year}{2018}).
\newblock


\bibitem[\protect\citeauthoryear{Heyburn, Bond, Black, Mulvenna, Wallace,
  Rankin, and Cleland}{Heyburn et~al\mbox{.}}{2018}]%
        {heyburn2018machine}
\bibfield{author}{\bibinfo{person}{Rachel Heyburn}, \bibinfo{person}{Raymond~R
  Bond}, \bibinfo{person}{Michaela Black}, \bibinfo{person}{Maurice Mulvenna},
  \bibinfo{person}{Jonathan Wallace}, \bibinfo{person}{Deborah Rankin}, {and}
  \bibinfo{person}{Brian Cleland}.} \bibinfo{year}{2018}\natexlab{}.
\newblock \showarticletitle{Machine learning using synthetic and real data:
  similarity of evaluation metrics for different healthcare datasets and for
  different algorithms}. In \bibinfo{booktitle}{\emph{Data Science and
  Knowledge Engineering for Sensing Decision Support: Proceedings of the 13th
  International FLINS Conference (FLINS 2018)}}. World Scientific,
  \bibinfo{pages}{1281--1291}.
\newblock


\bibitem[\protect\citeauthoryear{Huang, Oosterhuis, de~Rijke, and van
  Hoof}{Huang et~al\mbox{.}}{2020}]%
        {huang2020keeping}
\bibfield{author}{\bibinfo{person}{Jin Huang}, \bibinfo{person}{Harrie
  Oosterhuis}, \bibinfo{person}{Maarten de Rijke}, {and} \bibinfo{person}{Herke
  van Hoof}.} \bibinfo{year}{2020}\natexlab{}.
\newblock \showarticletitle{Keeping dataset biases out of the simulation: A
  debiased simulator for reinforcement learning based recommender systems}. In
  \bibinfo{booktitle}{\emph{Fourteenth ACM Conference on Recommender Systems}}.
  \bibinfo{pages}{190--199}.
\newblock


\bibitem[\protect\citeauthoryear{Inau, Sack, Waltemath, and Zeleke}{Inau
  et~al\mbox{.}}{2021}]%
        {Inau2021Initiatives}
\bibfield{author}{\bibinfo{person}{Esther~Thea Inau}, \bibinfo{person}{Jean
  Sack}, \bibinfo{person}{Dagmar Waltemath}, {and}
  \bibinfo{person}{Atinkut~Alamirrew Zeleke}.} \bibinfo{year}{2021}\natexlab{}.
\newblock \showarticletitle{Initiatives, Concepts, and Implementation Practices
  of FAIR (Findable, Accessible, Interoperable, and Reusable) Data Principles
  in Health Data Stewardship Practice: Protocol for a Scoping Review}.
\newblock \bibinfo{journal}{\emph{JMIR Research Protocols}}
  \bibinfo{volume}{10}, \bibinfo{number}{2} (\bibinfo{year}{2021}),
  \bibinfo{pages}{e22505}.
\newblock
\showISSN{1929-0748}
\urldef\tempurl%
\url{https://doi.org/10.2196/22505}
\showDOI{\tempurl}


\bibitem[\protect\citeauthoryear{Joanna~Redden and Terzieva}{Joanna~Redden and
  Terzieva}{2020}]%
        {DataJustice}
\bibfield{author}{\bibinfo{person}{Jessica~Brand Joanna~Redden} {and}
  \bibinfo{person}{Vanesa Terzieva}.} \bibinfo{year}{2020}\natexlab{}.
\newblock \bibinfo{booktitle}{\emph{{D}ata {H}arm {R}ecord (Updated)}}.
\newblock
\newblock
\shownote{\url{https://datajusticelab.org/data-harm-record/}, Online; accessed
  22-August-2021.}


\bibitem[\protect\citeauthoryear{Jordon, Yoon, and van~der Schaar}{Jordon
  et~al\mbox{.}}{2018a}]%
        {jordon2018measuring}
\bibfield{author}{\bibinfo{person}{James Jordon}, \bibinfo{person}{Jinsung
  Yoon}, {and} \bibinfo{person}{Mihaela van~der Schaar}.}
  \bibinfo{year}{2018}\natexlab{a}.
\newblock \showarticletitle{Measuring the quality of synthetic data for use in
  competitions}.
\newblock \bibinfo{journal}{\emph{arXiv preprint arXiv:1806.11345}}
  (\bibinfo{year}{2018}).
\newblock


\bibitem[\protect\citeauthoryear{Jordon, Yoon, and Van Der~Schaar}{Jordon
  et~al\mbox{.}}{2018b}]%
        {jordon2018pate}
\bibfield{author}{\bibinfo{person}{James Jordon}, \bibinfo{person}{Jinsung
  Yoon}, {and} \bibinfo{person}{Mihaela Van Der~Schaar}.}
  \bibinfo{year}{2018}\natexlab{b}.
\newblock \showarticletitle{PATE-GAN: Generating synthetic data with
  differential privacy guarantees}. In \bibinfo{booktitle}{\emph{International
  conference on learning representations}}.
\newblock


\bibitem[\protect\citeauthoryear{Krishnan, Patel, Franklin, and
  Goldberg}{Krishnan et~al\mbox{.}}{2014}]%
        {krishnan2014methodology}
\bibfield{author}{\bibinfo{person}{Sanjay Krishnan}, \bibinfo{person}{Jay
  Patel}, \bibinfo{person}{Michael~J Franklin}, {and} \bibinfo{person}{Ken
  Goldberg}.} \bibinfo{year}{2014}\natexlab{}.
\newblock \showarticletitle{A methodology for learning, analyzing, and
  mitigating social influence bias in recommender systems}. In
  \bibinfo{booktitle}{\emph{Proceedings of the 8th ACM Conference on
  Recommender systems}}. \bibinfo{pages}{137--144}.
\newblock


\bibitem[\protect\citeauthoryear{Krishnaraj}{Krishnaraj}{2019}]%
        {krishnaraj2019comparative}
\bibfield{author}{\bibinfo{person}{Manoj Krishnaraj}.}
  \bibinfo{year}{2019}\natexlab{}.
\newblock \showarticletitle{Comparative Analysis of Techniques for Data
  Minimization for Recommender System algorithms}.
\newblock  (\bibinfo{year}{2019}).
\newblock


\bibitem[\protect\citeauthoryear{Larson and Slokom}{Larson and Slokom}{2019}]%
        {larson2019up}
\bibfield{author}{\bibinfo{person}{Martha Larson} {and} \bibinfo{person}{Manel
  Slokom}.} \bibinfo{year}{2019}\natexlab{}.
\newblock \showarticletitle{Up close, but not too personal: Hypotargeting for
  recommender systems}. In \bibinfo{booktitle}{\emph{ImpactRS 2019 Workshop, in
  conjunction with the 11th ACM Conference on Recommender Systems (RecSys)}}.
\newblock


\bibitem[\protect\citeauthoryear{Larson, Zito, Loni, and Cremonesi}{Larson
  et~al\mbox{.}}{2017}]%
        {larson2017towards}
\bibfield{author}{\bibinfo{person}{Martha Larson}, \bibinfo{person}{Alessandro
  Zito}, \bibinfo{person}{Babak Loni}, {and} \bibinfo{person}{Paolo
  Cremonesi}.} \bibinfo{year}{2017}\natexlab{}.
\newblock \showarticletitle{Towards minimal necessary data: The case for
  analyzing training data requirements of recommender algorithms}. In
  \bibinfo{booktitle}{\emph{FATREC 2017 Workshop on Fairness, Accountability,
  and Transparency in Recommender Systems, in conjunction with the 11th ACM
  Conference on Recommender Systems (RecSys)}}.
\newblock


\bibitem[\protect\citeauthoryear{Li, Chen, Fu, Ge, and Zhang}{Li
  et~al\mbox{.}}{2021}]%
        {li2021user}
\bibfield{author}{\bibinfo{person}{Yunqi Li}, \bibinfo{person}{Hanxiong Chen},
  \bibinfo{person}{Zuohui Fu}, \bibinfo{person}{Yingqiang Ge}, {and}
  \bibinfo{person}{Yongfeng Zhang}.} \bibinfo{year}{2021}\natexlab{}.
\newblock \showarticletitle{User-oriented Fairness in Recommendation}. In
  \bibinfo{booktitle}{\emph{Proceedings of the Web Conference}}.
  \bibinfo{pages}{624--632}.
\newblock


\bibitem[\protect\citeauthoryear{Lin, Sonboli, Mobasher, and Burke}{Lin
  et~al\mbox{.}}{2019}]%
        {lin2019crank}
\bibfield{author}{\bibinfo{person}{Kun Lin}, \bibinfo{person}{Nasim Sonboli},
  \bibinfo{person}{Bamshad Mobasher}, {and} \bibinfo{person}{Robin Burke}.}
  \bibinfo{year}{2019}\natexlab{}.
\newblock \showarticletitle{Crank up the volume: preference bias amplification
  in collaborative recommendation}. In \bibinfo{booktitle}{\emph{in
  Multi-stakeholder Environments (RMSE’19), in conjunction with the 13th ACM
  Conference on Recommender Systems (RecSys’19)}}.
\newblock


\bibitem[\protect\citeauthoryear{Mansoury, Mobasher, Burke, and
  Pechenizkiy}{Mansoury et~al\mbox{.}}{2019}]%
        {mansoury2019bias}
\bibfield{author}{\bibinfo{person}{Masoud Mansoury}, \bibinfo{person}{Bamshad
  Mobasher}, \bibinfo{person}{Robin Burke}, {and} \bibinfo{person}{Mykola
  Pechenizkiy}.} \bibinfo{year}{2019}\natexlab{}.
\newblock \showarticletitle{Bias Disparity in Collaborative Recommendation:
  Algorithmic Evaluation and Comparison}. In \bibinfo{booktitle}{\emph{in
  Multi-stakeholder Environments (RMSE’19), in conjunction with the 13th ACM
  Conference on Recommender Systems (RecSys’19)}}.
\newblock


\bibitem[\protect\citeauthoryear{Narayanan and Shmatikov}{Narayanan and
  Shmatikov}{2008}]%
        {narayanan2008robust}
\bibfield{author}{\bibinfo{person}{Arvind Narayanan} {and}
  \bibinfo{person}{Vitaly Shmatikov}.} \bibinfo{year}{2008}\natexlab{}.
\newblock \showarticletitle{Robust de-anonymization of large sparse datasets}.
  In \bibinfo{booktitle}{\emph{IEEE Symposium on Security and Privacy}}.
  \bibinfo{pages}{111--125}.
\newblock


\bibitem[\protect\citeauthoryear{OpenAIRE}{OpenAIRE}{2018}]%
        {OpenAIRE}
\bibfield{author}{\bibinfo{person}{OpenAIRE}.} \bibinfo{year}{2018}\natexlab{}.
\newblock \bibinfo{booktitle}{\emph{The Four Basics of {FAIR}}}.
\newblock
\newblock
\shownote{\url{https://www.openaire.eu/what-is-fair-data}, Online; accessed
  22-August-2021.}


\bibitem[\protect\citeauthoryear{Regulation}{Regulation}{2018}]%
        {regulation2018principles}
\bibfield{author}{\bibinfo{person}{General Data~Protection Regulation}.}
  \bibinfo{year}{2018}\natexlab{}.
\newblock \bibinfo{title}{Principles Relating to Processing of Personal Data}.
\newblock
\newblock


\bibitem[\protect\citeauthoryear{Said and Bellog\'{\i}n}{Said and
  Bellog\'{\i}n}{2014}]%
        {saidbellogin2014}
\bibfield{author}{\bibinfo{person}{Alan Said} {and} \bibinfo{person}{Alejandro
  Bellog\'{\i}n}.} \bibinfo{year}{2014}\natexlab{}.
\newblock \showarticletitle{Comparative Recommender System Evaluation:
  Benchmarking Recommendation Frameworks}. In
  \bibinfo{booktitle}{\emph{Proceedings of the 8th ACM Conference on
  Recommender Systems}} (Foster City, Silicon Valley, California, USA)
  \emph{(\bibinfo{series}{RecSys '14})}. \bibinfo{publisher}{Association for
  Computing Machinery}, \bibinfo{address}{New York, NY, USA},
  \bibinfo{pages}{129–136}.
\newblock
\showISBNx{9781450326681}
\urldef\tempurl%
\url{https://doi.org/10.1145/2645710.2645746}
\showDOI{\tempurl}


\bibitem[\protect\citeauthoryear{Slokom, Hanjalic, and Larson}{Slokom
  et~al\mbox{.}}{2021}]%
        {slokom2021PerBlur}
\bibfield{author}{\bibinfo{person}{Manel Slokom}, \bibinfo{person}{Alan
  Hanjalic}, {and} \bibinfo{person}{Martha Larson}.}
  \bibinfo{year}{2021}\natexlab{}.
\newblock \showarticletitle{Towards user-oriented privacy for recommender
  system data: A personalization-based approach to gender obfuscation for user
  profiles}.
\newblock \bibinfo{journal}{\emph{Information Processing \& Management}}
  \bibinfo{volume}{58}, \bibinfo{number}{6} (\bibinfo{year}{2021}),
  \bibinfo{pages}{102722}.
\newblock
\showISSN{0306-4573}
\urldef\tempurl%
\url{https://doi.org/10.1016/j.ipm.2021.102722}
\showDOI{\tempurl}


\bibitem[\protect\citeauthoryear{Slokom, Larson, and Hanjalic}{Slokom
  et~al\mbox{.}}{2019}]%
        {slokom2019data}
\bibfield{author}{\bibinfo{person}{Manel Slokom}, \bibinfo{person}{Martha
  Larson}, {and} \bibinfo{person}{Alan Hanjalic}.}
  \bibinfo{year}{2019}\natexlab{}.
\newblock \showarticletitle{Data Masking for Recommender Systems: Prediction
  Performance and Rating Hiding}.
\newblock \bibinfo{journal}{\emph{Late breaking results, in conjunction with
  the 13th ACM Conference on Recommender Systems, RecSys}}
  (\bibinfo{year}{2019}).
\newblock


\bibitem[\protect\citeauthoryear{Tian and Ekstrand}{Tian and Ekstrand}{2018}]%
        {tian2018monte}
\bibfield{author}{\bibinfo{person}{Mucun Tian} {and} \bibinfo{person}{Michael~D
  Ekstrand}.} \bibinfo{year}{2018}\natexlab{}.
\newblock \showarticletitle{Monte Carlo Estimates of Evaluation Metric Error
  and Bias}. In \bibinfo{booktitle}{\emph{REVEAL 2018 Workshop on Offline
  Evaluation in Recommender Systems, in conjunction with the 12th ACM
  Conference on Recommender Systems (RecSys’18)}}.
\newblock


\bibitem[\protect\citeauthoryear{Tsintzou, Pitoura, and Tsaparas}{Tsintzou
  et~al\mbox{.}}{2018}]%
        {tsintzou2018bias}
\bibfield{author}{\bibinfo{person}{Virginia Tsintzou},
  \bibinfo{person}{Evaggelia Pitoura}, {and} \bibinfo{person}{Panayiotis
  Tsaparas}.} \bibinfo{year}{2018}\natexlab{}.
\newblock \showarticletitle{Bias Disparity in Recommendation Systems}.
\newblock \bibinfo{journal}{\emph{arXiv preprint arXiv:1811.01461}}
  (\bibinfo{year}{2018}).
\newblock


\bibitem[\protect\citeauthoryear{Yao, Halpern, Thain, Wang, Lee, Prost, Chi,
  Chen, and Beutel}{Yao et~al\mbox{.}}{2021}]%
        {yao2021measuring}
\bibfield{author}{\bibinfo{person}{Sirui Yao}, \bibinfo{person}{Yoni Halpern},
  \bibinfo{person}{Nithum Thain}, \bibinfo{person}{Xuezhi Wang},
  \bibinfo{person}{Kang Lee}, \bibinfo{person}{Flavien Prost},
  \bibinfo{person}{Ed~H Chi}, \bibinfo{person}{Jilin Chen}, {and}
  \bibinfo{person}{Alex Beutel}.} \bibinfo{year}{2021}\natexlab{}.
\newblock \showarticletitle{Measuring Recommender System Effects with Simulated
  Users}.
\newblock \bibinfo{journal}{\emph{arXiv preprint arXiv:2101.04526}}
  (\bibinfo{year}{2021}).
\newblock


\end{thebibliography}



\end{document}